\documentclass[fleqn,10pt]{wlscirep}

\newcommand{\mps}[1]{Mn$_{#1}$PtSn}

\newcommand{\tc}{\textit{T}\textsubscript{C}}
\newcommand{\ts}{\textit{T}\textsubscript{s}}
\newcommand{\rt}{$\rho_{\mathrm{xy}}^{\mathrm{THE}}$}
\newcommand{\ra}{$\rho_{\mathrm{xy}}^{\mathrm{AHE}}$}
\newcommand{\ro}{$\rho_{\mathrm{xy}}^{\mathrm{OHE}}$}
\newcommand{\rtm}{$\rho_{\mathrm{xy}}^{\mathrm{THE,max}}$}

\newcommand{\dmdh}{$\mathrm{d}M/\mathrm{d}H$}
\newcommand{\bnc}{$\mu_{0}H_{\mathrm{nc}}$}

\newcommand{\reffilm}{110~nm thick Mn$_{1.61}$PtSn}
\newcommand{\rxy}{$\rho$\textsubscript{xy}}
\newcommand{\conc}[1]{Mn($x$)=#1}
\newcommand{\concx}{Mn($x$)}
\newcommand{\hs}{$\mu_{0}H_{\mathrm{s}}$}

\usepackage{graphicx}
\usepackage[utf8]{inputenc}
\usepackage{hyperref}
\usepackage{verbatim}
\usepackage{gensymb}
\usepackage{mcite}
\usepackage{upgreek}

\makeatletter
\renewcommand{\fnum@figure}{Fig. \thefigure}
\makeatother

\begin{document}
	
	\title{Role of magnetic exchange interactions in chiral-type Hall effects of epitaxial {Mn$_{x}$PtSn} films}
	
	\author[1,2]{Peter Swekis}
	\author[1]{Jacob Gayles}
	\author[1,3]{Dominik Kriegner}
	\author[1]{Gerhard H. Fecher}
	\author[1]{Yan Sun}	
	\author[2]{Sebastian T. B. Goennenwein}
	\author[1]{Claudia Felser}
	\author[1]{Anastasios Markou}
	
	\affil[1]{Max Planck Institute for Chemical Physics of Solids, 01187 Dresden, Germany}
	\affil[2]{Institut f{\"u}r Festk{\"o}rper- und Materialphysik, Technische Universit{\"a}t Dresden, 01062 Dresden, Germany}
	\affil[3]{Institute of Physics ASCR, v.v.i., Cukrovarnicka 10, 162 53, Praha 6, Czech Republic}

	\affil[*]{anastasios.markou@cpfs.mpg.de}
	\affil[*]{claudia.felser@cpfs.mpg.de}
	
	\date{\today}
	
	\begin{abstract}
	 Tetragonal Mn-based Heusler compounds feature rich exchange interactions and exotic topological magnetic textures, such as antiskyrmions, complimented by the chiral-type Hall effects. This makes the material class interesting for device applications. We report the relation of the magnetic exchange interactions to the thickness and Mn concentration of \mps{x}~thin films, grown by magnetron sputtering. The competition of the magnetic exchange interactions determines the finite temperature magnetic texture and thereby the chiral-type Hall effects in external magnetic fields. We investigate the magnetic and transport properties as a function of magnetic field and temperature. We focus on the anomalous and chiral-type Hall effects and the behavior of the dc-magnetization, in relation to chiral spin textures. We further determine the stable crystal phase for a relative Mn concentration between 1.5 and 1.85 in the $I\overline{4}2d$ structure. We observe a spin-reorientation transition in all compounds studied, which is due to the competition of exchange interactions on different Mn sublattices. We discuss our results in terms of exchange interactions and compare them with theoretical atomistic spin calculations.	
	\end{abstract}
	
	\maketitle
	
\section*{Introduction}
Magnetic spin textures are of vital interest for technological applications due to the ease of manipulation via external electromagnetic fields~\cite{Fert2013,*Parkin2015,Caretta2018} and detection purely through electrical means~\cite{Bruno2004}. One such method for the detection for complex spin textures arises from the Hall effects, i.e. transverse electric fields arise in response to an applied external electric field. The well-known ordinary (OHE) and anomalous (AHE) Hall effects scale with external magnetic field and the alignment of magnetic domains in the absence of a magnetic field, respectively~\cite{Nagaosa2010}. In addition, there are transport effects that are independent of the external magnetic field and net magnetization, which we group into the chiral-type Hall effects~\cite{Onoda2004,Lux2019,Denisov2018,Rozhansky2019}. These chiral-type Hall effects are experimentally determined by the subtraction of the AHE and OHE from the total Hall effect~\cite{Neubauer2009} and are related to the spatial variation of the local magnetic lattice in real space~\cite{Bruno2004}. The relative spatial variation gives rise to an emergent electromagnetic field, termed the Berry curvature, which is finite for localized magnetic textures with spin chirality. Electromagnetic fields couple to those spin textures through the conduction electrons and in turn the ground state electronic properties~\cite{Grytsiuk2020}.
	
The ground state electronic properties determine the magnetic exchange interactions necessary to stabilize spin textures at finite temperatures and magnetic fields. These exchange interactions arise in a hierarchical competition of energy scales: the symmetric Heisenberg exchange which is nominally the largest and determines the ordering temperature; the anti-symmetric Dzyaloshinskii-Moriya interaction (DMI) attributed to the absence of inversion symmetry and strong spin-orbit coupling (SOC)~\cite{Dzyaloshinsky1958,*Moriya1960}; the dipole-dipole interactions; the magnetocrystalline anisotropy. The competition of these interactions can lead to magnetic textures that are noncoplanar with a finite chiral Berry curvature. One example is the skyrmion crystal arising in a competition of the Heisenberg exchange with the DMI, found in B20 compounds~\cite{Neubauer2009,Muehlbauer2009,Wilson2012,Wilson2014,Kanazawa2011,*Huang2012,*Franz2014,*Yokouchi2014,*Spencer2018}, or the antiskyrmion in inverse tetragonal Heusler compounds~\cite{Nayak2017,*Jena2019,*Saha2019}. The Hall effect from such magnetic textures is nominally the topological Hall effect (THE) since it can be directly related to the topological winding of the skyrmion lattice~\cite{Bruno2004}. Systems with electron correlation can also show this type of response, which can sometimes be associated with magnetic domains and bubbles~\cite{Ohuchi2015,*Takahashi2018,*Vistoli2018,*Nakamura2018,*Qin2019}. Lastly, there are the noncollinear spin arrangements that can originate in the competition of ferromagnetic and antiferromagnetic exchange interactions between neighboring spins~\cite{Suergers2014,*Liu2017,*Rout2019,*Yan2019,*Taylor2019}.
	
The class of tetragonal inverse Mn-based Heusler compounds ($Mn$$_{x}YZ$, $Y$ is a transition metal and $Z$ a main-group element), associated with a broken inversion symmetry ($D_{2d}$) that enables the stabilization of noncoplanar spin textures, received significant attention following the observation of antiskyrmions with an anisotropic winding over a broad temperature and field range~\cite{Nayak2017,*Saha2019,*Jena2019,Vir2019a}. These multi-component systems with several magnetic sublattices are particularly interesting, in that the magnetic interactions depend on the chemical substitution~\cite{Graf2011,Wollmann2015}. The SOC can be changed by partial substitution of the main group element $Z$ and the $5d$-transition metal $Y$, effectively tuning the DMI and, in turn, the spin texture (e.g., antiskyrmion phase)~\cite{Kumar2020}. The Mn concentration has a profound effect on the competition of the Heisenberg exchange interactions~\cite{Winterlik2012}, which impact the canting strength between the two distinct Mn sublattices and, in turn, the temperature of the spin reorientation from a collinear into a noncollinear spin texture~\cite{Meshcheriakova2014,Herran2018}. Simultaneously, the Mn concentration overns to the magnetocrystalline anisotropy~\cite{Winterlik2012} and the large moment of Mn induces a strong dipole-dipole interaction. Confined geometries and tunable anisotropies~\cite{Srivastava2020}, accessible in thin films~\cite{Wilson2012,Wilson2014} play another crucial role for the stabilization and control of those complex magnetic textures. In that context, thin films allow the design and engineering of devices on technologically relevant scales that are easily integratabtle into existing concepts.
	
In this work, we experimentally investigate thin films of the compound \mps{x}, grown by magnetron sputtering, as a function of the overall Mn concentration ($1.48\le x\le2.04$) and film thickness (20~nm$\le t \le$110~nm). The Hall effects, measured on micro-patterned devices, are extracted in a broad temperature and applied magnetic field range. We capture the change in the magnetic structure via the dc-susceptibility as a function of the external magnetic field strength. Furthermore, we relate the transition temperatures to a trend in exchange interactions. First principle calculations confirm a change in the ground state exchange interactions as a function of Mn concentration. We find that a significant contribution to the Hall resistivity originates in the chiral spin structure below the spin reorientation, which can be tuned by the Mn concentration. Furthermore, the exchange that gives rise to the chiral spin structure can act as an additional source to the magnetic anisotropy, which is modeled by atomistic spin simulations. Finally, we compare the observed topological Hall effect to the anomalous Hall effect as a function of film thickness and Mn concentration, and find that they are of comparable size in the thin-film limit.
	
\section*{Materials and methods}
	
\noindent{\fontfamily{phv}\selectfont\textbf{Thin film growth.}} Epitaxial thin films of \mps{x}~were grown in a BESTEC UHV magnetron sputtering system on single crystal MgO (001) substrates as described in ref.\cite{Swekis2019}. The constituting elements (Mn, Pt, Sn) were deposited from 5.1~cm targets by dc magnetron co-sputtering. Here, the stoichiometry was controlled by adjusting the power of the sources. The deposition was performed in confocal geometry with a target to substrate distance of 20~cm. Prior to deposition, the chamber was evacuated to a base pressure below $2\times 10^{-8}$~mbar, while during deposition a process gas pressure of $3\times 10^{-3}$~mbar (Ar, 15~sccm) was maintained. The films were deposited at $350$~$\degree$C and post-annealed for 30~min at the same temperature in order to improve the chemical ordering. The annealed films were capped at room temperature with 3~nm Si, in order to prevent oxidation.
	
\noindent{\fontfamily{phv}\selectfont\textbf{Crystal structure and stoichiometry.}} The stoichiometries of the films were confirmed by energy-dispersive x-ray spectroscopy (EDXS), with an experimental uncertainty of 5~at.\%. Structural characterization was carried out on a PANalytical X’Pert PRO x-ray diffractometry (XRD) system employing  Cu-K$\alpha_1$ radiation ($\lambda =1.5406$ \AA). The film thicknesses were determined by x-ray reflectivity (XRR) measurements.
	
\noindent{\fontfamily{phv}\selectfont\textbf{Hard x-ray photoelectron spectroscopy.}} The experiments were performed at beamline P22~\cite{Schlueter2019} of PETRA~III (Hamburg, Germany) using linearly polarized photons with an energy of $h\nu=5226.1$~eV for excitation. Horizontal ($p$) polarization was obtained directly from the undulator. Grazing incidence ($\alpha=89^\circ$) -- normal emission ($\theta=1^\circ$) geometry was used ensuring a polarization vector nearly parallel to the surface normal. The energy resolution was set to 250~meV, verified by spectra of the Au valence band at the Fermi energy \textit{$E$}$_{\mathrm F}$. The spectra were acquired using a Phoibos 225HV photoelectron spectrometer with a hemispherical analyzer, delayline detector and wide angle lens (SPECS, Berlin). The spectra are not influenced by the 3~nm thick Si capping layer, due to the large electron mean free path and low cross section of its valence states~\cite{Fecher2008}.
	
\noindent{\fontfamily{phv}\selectfont\textbf{DC magnetometry measurements.}} Magnetization measurements were performed on a SQUID vibrating sample magnetometer (MPMS 3, Quantum Design). In order to infer the magnetization of the films, we subtracted the diamagnetic substrate contribution as well as a low-temperature paramagnetic contribution from the raw data. Here, the paramagnetic contribution is attributed to impurities in the MgO substrate. The diamagnetic susceptibility of MgO was estimated from the high field slope at 300~K. The paramagnetic contribution was fitted and subtracted from the raw data using the Brillouin function.
	
\noindent{\fontfamily{phv}\selectfont\textbf{Lithography and electrical transport measurements.}} In order to determine the longitudinal and the Hall resistivity, the thin films were patterned into 8-contact Hall-bar devices (width = 50~$\upmu$m; length = 250~$\upmu$m) using optical lithography and subsequent Ar ion etching (Fig.~\ref{fig:F4}b). The electrical transport measurements were conducted on a physical properties measurement system (PPMS, Quantum Design) with fields up to 7~T applied along the out-of-plane (z) direction (MgO [001]). An in-plane current, $I_{\mathrm{x}}$=1~mA (along MgO[100]), was applied along the Hall-bar while the voltages along the current direction ($V_{\mathrm{xx}}$, x-direction corresponding to MgO[100]) and perpendicular to the current direction ($V_{\mathrm{xy}}$, y-direction corresponding to MgO[010]) were recorded simultaneously.

\noindent{\fontfamily{phv}\selectfont\textbf{First principle calculations.}} We carried out density functional theory (DFT) calculations using the full potential linearized augmented plane-wave method within the generalized gradient approximation (GGA) exchange correlation (XC) functional. We converged the unit cell with 12x12x8 $k$-point mesh and 4.0~bohr$^{-1}$ cutoff radius for a spin full calculation. Spin spiral calculations were made within the generalized Bloch theorem formalism, with Anderson's force theorem. For the calculation of the AHE we used 64 Wannier functions with spin-orbit coupling for each unit cell~\cite{FLEUR}.
	
\section*{Results and discussion}
\subsection*{Crystal structure} 
In our previous work we reported the crystal structure of inverse tetragonal \mps{1.5}~thin films derived from the $I\overline{4}2d$ bulk structure, with a broken equivalence between the $a$ (out-of-plane) and $b$ (in-plane) lattice parameters, due to the film geometry, resulting in the $I2_{1}2_{1}2_{1}$ structure~\cite{Swekis2019}. Furthermore, the films grow in two distinct orientations, i.e., the tetragonal $c$-axis [001] lies either along the [110] or the [1$\overline{1}$0] axis of the MgO substrate~\cite{Swekis2019} (see Fig.~S1). In the following, we will discuss how the relative Mn concentration (\concx), as well as the thickness ($t$), affect the structural properties of those films.
	
\begin{figure}
	\centering
	\includegraphics[width=\columnwidth]{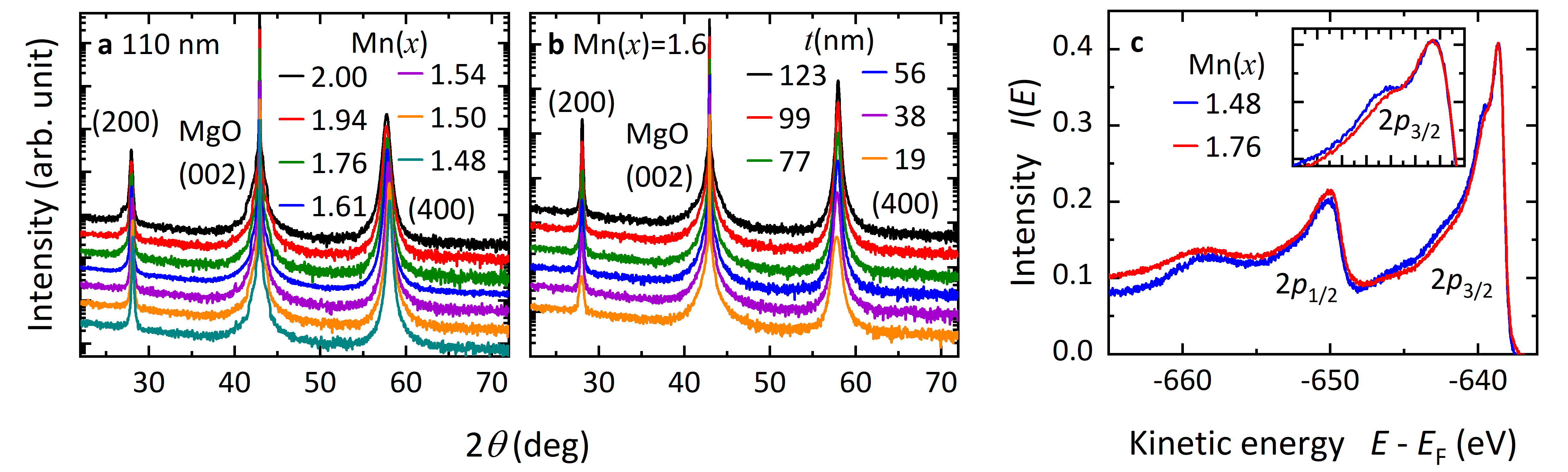}\\
	\caption{Symmetric radial ($\omega-2\theta$) x-ray diffractometry scans of \mps{x}~films recorded in out-of-plane geometry: (a) 110~nm thick films with different \concx~and (b) \conc{1.6}~films with different thickness. (c) Mn $2p$ photoelectron spectra of \conc{1.48}~and \conc{1.76}~films. Inset: Mn $2p_{3/2}$ derived peak.}
	\label{fig:F1}
\end{figure}	
	
Symmetric radial ($\omega-2\theta$) x-ray diffractometry (XRD) scans are employed to determine phase purity, verify the previously reported structure, and investigate changes in the lattice parameters. We show in Fig.~\ref{fig:F1}a the $\omega-2\theta$ scans of 110~nm thick \mps{x}~films at various \concx. Between \conc{1.5}~and \conc{1.85} all films grow epitaxially without any additional phases. In this range, we observe only the ($h$00) Bragg refections, attributed to the \mps{x}~film with the structure as mentioned above. Above \conc{1.85}~a shoulder forms at the (200) peak at around 27.0\degree, while below \conc{1.5}~additional peaks start to appear at around 40.2\degree~and 60.3\degree. The out-of-plane $a$-lattice parameter and the unit cell volume increase with increasing \concx, depending sensitively on the stoichiometry up to \conc{1.6}~ and followed by saturation towards \conc{1.85} (see Fig.~S3). Figure~\ref{fig:F1}b depicts the $\omega-2\theta$ scans of the film thickness dependence at \conc{1.6}. Here, we confirm phase purity and epitaxial growth for all thicknesses. The bulk values for the lattice parameters ($a=6.3651$~\AA, $c=12.2205$~\AA ~~for Mn$_{1.44}$PtSn~\cite{Vir2019b}) suggest a lattice mismatch with the MgO substrate of 6.4\% and 2.5\% along the two distinct in-plane directions of Mn$_x$PtSn. This results in an in-plane compressive strain and in turn an elongation of the out-of-plane lattice parameter ($a$). Upon relaxation of the compressive strain with increased $t$, the $a$-lattice parameters, therefore, decrease towards the bulk limit (see Fig.~S3). Further information on the structure can be found in the supplemental material.
	
\subsection*{Photoemission} 
We investigate the core levels and the valence bands of \mps{x}~films by hard x-ray photoelectron spectroscopy (HAXPES). An analysis of the peak heights of the Mn $2p$, Pt $4d$, and Sn $3d$ states (not shown here) confirms the trend of the relative \concx~between the films. Fig.~\ref{fig:F1}c compares the Mn $2p$ photoelectron spectra of \conc{1.48}~and \conc{1.76}. The $2p$ states exhibit a clear spin-orbit splitting of 12.4~eV, independent of the composition. Further, the spectra are dominated by an exchange splitting due to an interaction between the $2p$ core hole with the valence $3d$ states resulting in a pronounced multiplet spectrum, well known for Mn~\cite{Martins2006}. Since the Mn atoms carry a localized magnetic moment related to the magnetic splitting of the $3d$ states, the multiplet structure of the Mn $2p$ state reflects the magnetic state. It is seen from the inset in Fig.~\ref{fig:F1}c that the multiplet splitting at the Mn $2p_{3/2}$ derived peak is weaker at a higher \concx.
	
\subsection*{Magnetic spin reorientation}
\begin{figure}
	\centering
	\includegraphics[width=\textwidth]{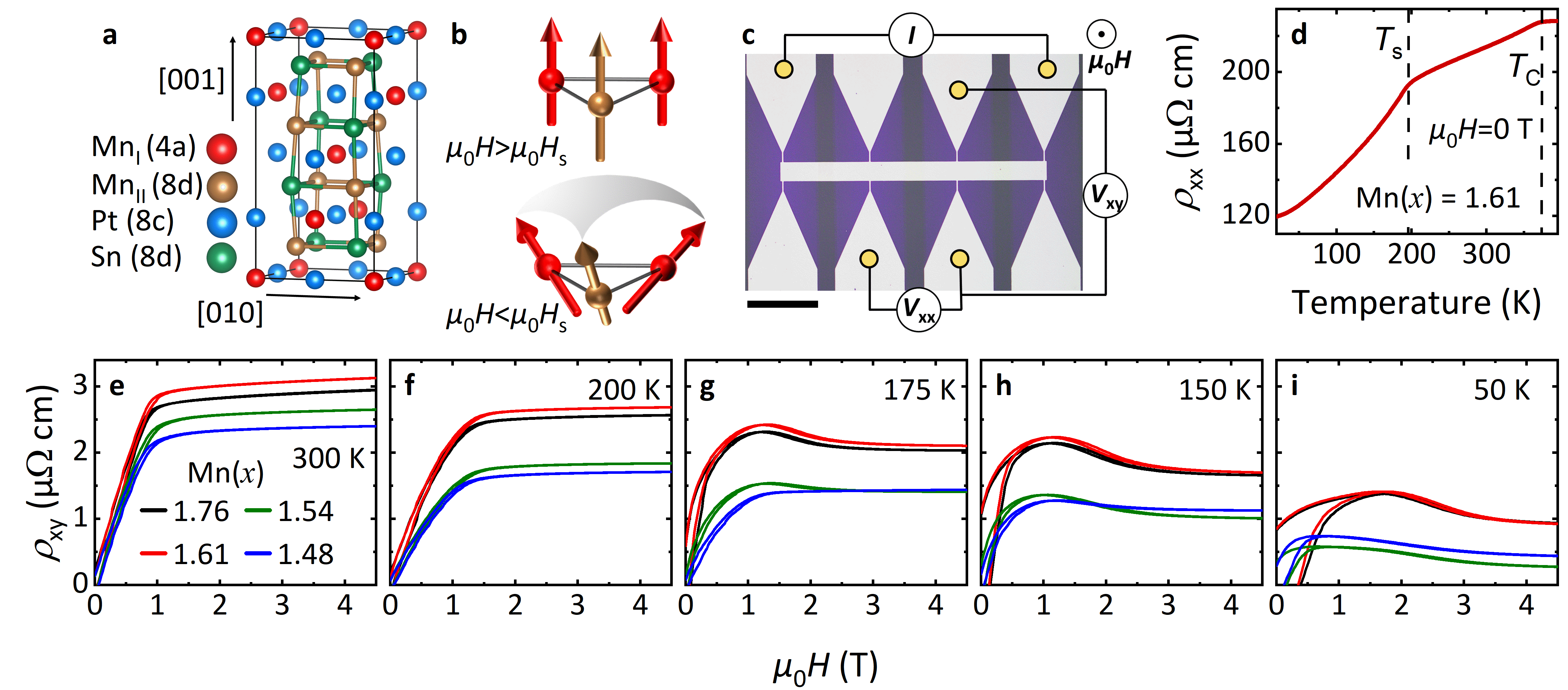}\\
	\caption{(a) Crystal structure of \mps{x}. (b) Magnetic spin texture prior and after field polarization at \hs. (c) Optical micrograph of the Hall-bar device. The black scalebar is 200 $\mu$m wide. (d) Zero field longitudinal resistivity of a \reffilm~film as a function of temperature. The dashed line at \ts$=195$~K indicates the spin reorientation transition temperature, while the dashed line at \tc$=377$~K indicates the Curie temperature. (e)-(i) 1\textsuperscript{st} quadrant of the measured Hall data for 110~nm thick \mps{x}~films with varying \concx~at 300~K, 200~K, 175~K, 150~K and 50~K, respectively.}
		\label{fig:F4}
\end{figure}

The magnetic structure of \mps{x} displays interesting magnetic transitions as a function of temperature and external field~\cite{Swekis2019,Vir2019a}. In the absence of an external magnetic field and as the temperature decreases from the paramagnetic state, there is a transition to a collinear state at the Curie temperature (\tc). Further decreasing the temperature leads to a first-order transition, from the collinear to a noncoplanar magnetic state, at the spin reorientation transition temperature (\ts)~\cite{Meshcheriakova2014}. In Fig.~\ref{fig:F4}b, we show a schematic of the spin reorientation as a function of an external magnetic field below \ts. There are two distinct magnetic sublattices of the Mn ions (Mn$_{\mathrm{I}}$-$4a$ and Mn$_{\mathrm{II}}$-$8d$) shown in gold and in red in Fig.~\ref{fig:F4}a-b. In the demagnetized state, the localized spins within the unit cell form a noncoplanar structure. As an external magnetic field is applied, the magnetic domains and the noncoplanar spin texture orient along the direction of the applied magnetic field. The increase of the external field allows for a spin reorientation at \hs, where the spins show a second-order transition from the noncoplanar arrangement into a collinear ferromagnetic alignment. These transitions are evident in the resultant transport phenomena in the presence of an electric field~\cite{Swekis2019} and in \dmdh.

\subsection*{Magnetotransport} 	
 In Fig.~\ref{fig:F4}c, we show the lithographically patterned 8-contact Hall-bar device. We measure the longitudinal resistivity ($\rho$\textsubscript{xx}) in the absence of an external field as a function of temperature (Fig.~\ref{fig:F4}d). Here, the slope changes at $T=195$~and 377~K indicate \ts~and \tc, respectively. The magnetic structure adds linearly to the other resistivity sources (e.g., phonons and impurities) due to {\it Matthiessen's} rule. Above \ts~the slope of $\rho$\textsubscript{xx}($T$) decreases as compared to the noncoplanar configuration, where the magnetoresistance decreases due to the collinear spin arrangement. The residual resistivity ratio ({\it RRR}) of our samples ranges from 1.58-2.79 as a function of thickness and \concx (see Fig.~S8). The chiral textures also further contributes to the Hall effects, due to the associated Berry curvature~\cite{Nagaosa2010}. Hence, the total Hall resistivity can be described as~\cite{Nagaosa2010,Swekis2019,Schlitz2019},
	
\begin{equation}\label{eq:Hall}
\centering
	\rho_{xy}=\rho_{xy}^{\mathrm{OHE}}+\rho_{xy}^{\mathrm{AHE}}+\rho_{xy}^{\mathrm{THE}}.
\end{equation}
	
Equation (1) comprises, first, the ordinary Hall resistivity (\ro$=R_0H$) with the scaling parameter $R_0$ as the Hall constant, depending on the carrier density. Secondly, the anomalous Hall resistivity (\ra$=R_{\mathrm{S}}M$) associated with the spontaneous magnetization and the scaling parameter $R_{\mathrm{S}}$ that originates from the scattering mechanisms and the intrinsic Berry curvature. Lastly, the topological Hall resistivity (\rt) that originates from the chiral magnetic texture and is independent of the net magnetization. Therefore, it is straight-forward to extract \rt~by subtracting the contributions to \rxy~that scale with the external field and the saturation of magnetization~\cite{Neubauer2009,Nagaosa2010,Swekis2019} (see Fig.~S9).
	
We show the total Hall resistivities in Fig.~\ref{fig:F4}e-i measured on the micro-patterned devices under the application of an external magnetic field and as a function of \concx~and temperature. At 300~K and 200~K, \rxy~portrays the typical behavior for a ferromagnet following a saturating magnetization in an external magnetic field (Fig.~\ref{fig:F4}e-f). We determine \ra~as the linear extrapolation of the resistivity above 2.5~T to zero field. The size of \ra~varies with concentration related to the electron occupation determining the Berry curvature and the sources of scattering centers (see Fig.~S10). In the graphs, we observe the largest \ra~for \conc{1.61}.  Upon further decreasing the temperature (Fig.~\ref{fig:F4}g-i), we see a nonlinear, {\it hump-like}, behavior in \rxy~in the range of 0.5-2.5 T and a more considerable hysteresis~\cite{Swekis2019,Vir2019a,Kumar2020}. At 175~K, in Fig.~\ref{fig:F4}g, the {\it hump-like} feature is seen for all \concx~except \conc{1.48}, where it appears at 150~K. The overall magnitude of \ra~decreases with the decrease in temperature, as expected from the reduction in scattering mechanisms. For \rt~the following trend is apparent: zero above \ts; finite around \ts; maximal for temperatures below \ts. Further, there is a unique dependence of \rt~on \concx, which can only point to the magnetic interactions that determine the magnetic ordering of the system.

\begin{figure}[tp!]
	\centering
	\includegraphics[width=\columnwidth]{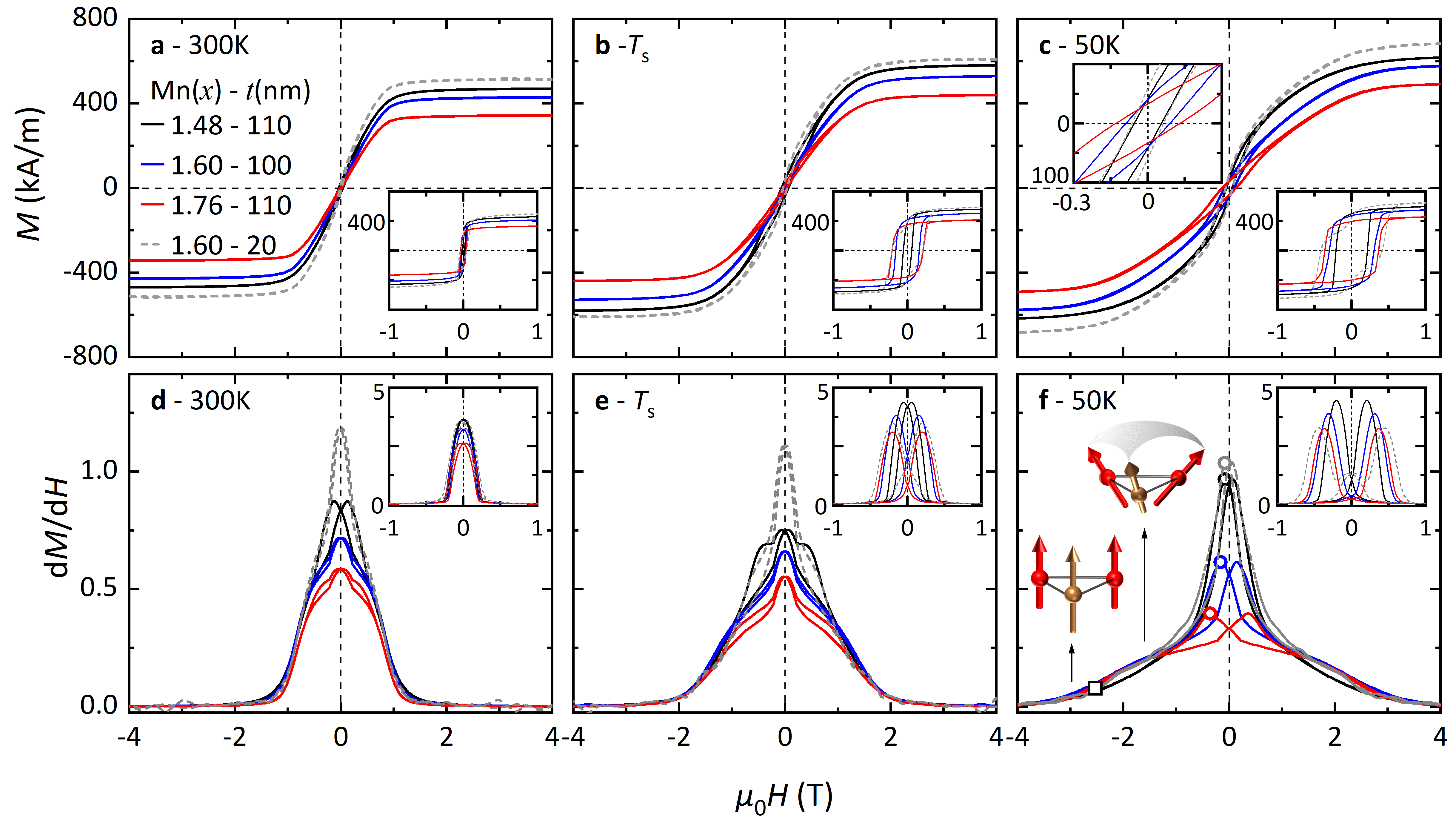}\\
	\caption{(a)-(c) $M(H)$ loops and (d)-(f) magnetic field dependence of \dmdh~calculated from the dc-magnetization. The field is applied along the film normal at 300~K, \ts~and 50~K, respectively, for \conc{1.48} and \conc{1.71} films from the 110~nm \concx~series and $t=20$ and 100~nm films from the \conc{1.6} thickness series. Squares and circles in (f) indicate the magnetic transitions. Insets right: $M(H)$ loops and \dmdh~with the field applied in the film plane. Inset top (c): Enlarged area around 0~T. Inset left (f): Magnetic spin texture below and above \hs.}	
	\label{fig:F2}
\end{figure}
	
\subsection*{DC magnetic susceptibility} 	
Changes in the magnetic texture can be inferred from the static magnetic susceptibility (\dmdh)~\cite{Bauer2012,Wilson2012,Wilson2014}. Hence, we perform dc-magnetometry measurements in order to extract the magnetization of the unpatterned samples. In Fig.~\ref{fig:F2}a-c we show the out-of-plane (OOP) magnetizations as a function of an externally applied field ($M(H)$ loops) at 300~K, \ts~and 50~K. The insets on the lower right are the in-plane (IP) $M(H)$ loops for the respective films. We show two samples with \conc{1.48} and \conc{1.71} from the 110~nm \concx~series as well as two samples at $t=20$ and 100~nm from the \conc{1.6} thickness series. The upper right inset of Fig.~\ref{fig:F2}c depicts a small hysteresis pocket around 0~T. The OOP $M(H)$ loops are reminiscent of hard-axis behavior with a small coercive field. The IP $M(H)$ loops display that the magnetic easy-axis lies within the film plane due to the anisotropy of the system and in agreement with the tetragonal $c$-axis orientation. Both, the coercive fields and the saturation magnetization increase with decreasing temperature.

In Fig.~\ref{fig:F2}d-f we plot the OOP \dmdh~as a function of the externally applied field. The static magnetic susceptibility is calculated from the the numerical derivative of the the experimental magnetization. The insets show the IP \dmdh. At 300~K (Fig.~\ref{fig:F2}d), the demagnetized textures rapidly converge to a saturated state at increasing fields of about 1~T. Below \ts~(Fig.~\ref{fig:F2}f) there are two transitions, close to 0~T (open circles) and $\sim$2.5~T (open square), both determined from peaks in the second derivative of the magnetization with respect to the external magnetic field. In this range, the magnetic domains begin to align along the field direction. The noncoplanar magnetic texture is dominant where \dmdh~slope is decreased compared to the slope close to 0~T. For fields larger than \hs, the magnetic structure is collinear and \dmdh~is a constant zero. The IP \dmdh~(Fig.~\ref{fig:F2}f (top inset)) shows two peaks that are symmetric around 0~T, which is due to the square loop hysteresis. In the spin reorientation(Fig.~\ref{fig:F2}e), \dmdh~is a mixture of the two states above and below \ts. Further, the hysteresis displayed in the inset of Fig.~\ref{fig:F2}c, does not belong to the magneto-crystalline anisotropy, but to the energy required to orient noncoplanar spins with respect to one another~\cite{Wilson2012,Jamaluddin2019}. This is a first-order transition highlighted by the circles close to 0~T (\bnc). The smooth second-ordered transition from the noncoplanar to the collinear spin arrangement is shown as schematic insets (Fig.~\ref{fig:F2}f). These effects originate in the magnetic exchange interactions, specifically, the competition between antiferromagnetic and ferromagnetic interactions on the Mn sublattices.

\begin{figure}
	\centering
	\includegraphics[width=\columnwidth]{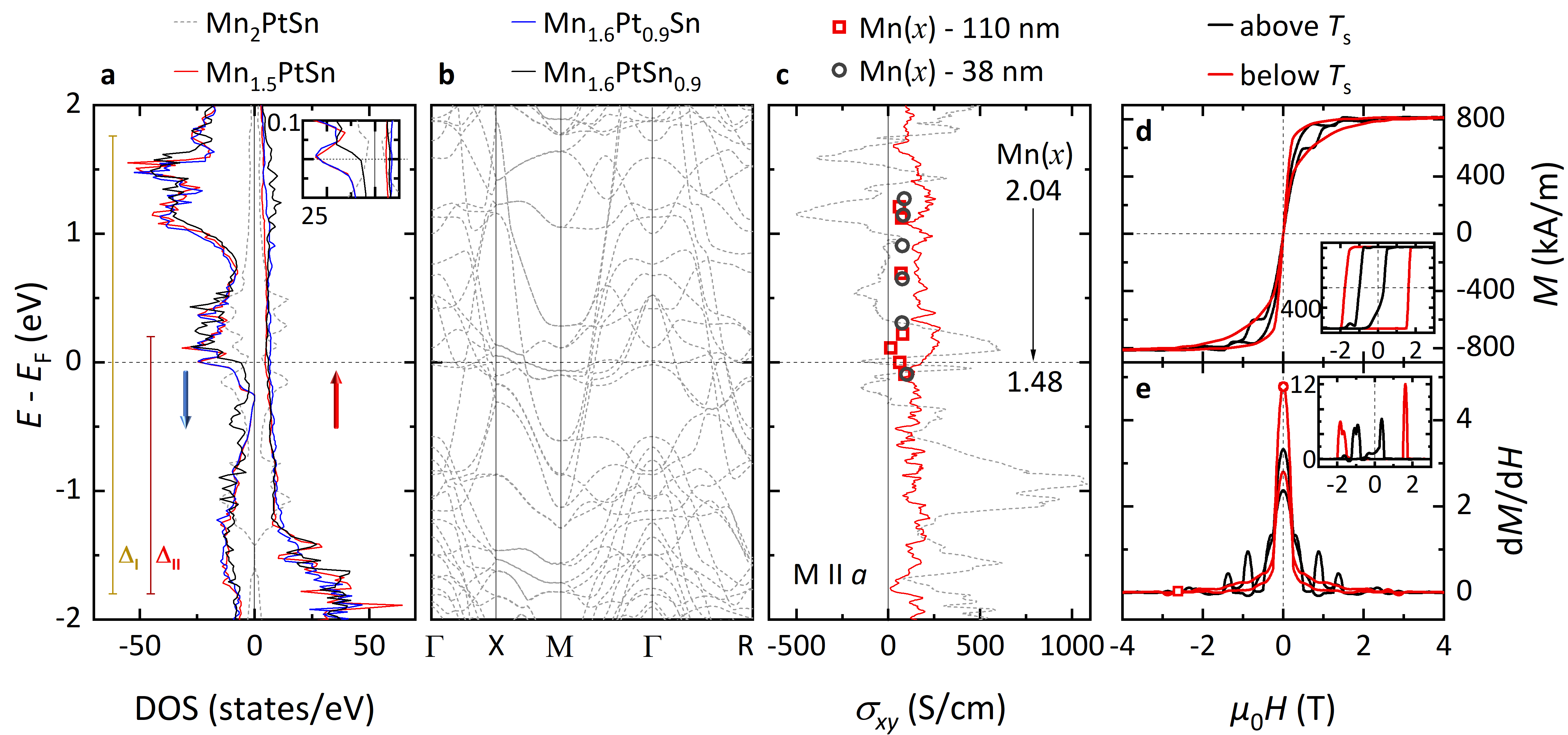}
	\caption{(a) Density of states as a function of the Fermi energy for \mps{2} (grey dashed), \mps{1.5} (solid red), Mn$_{1.6}$Pt$_{0.9}$Sn (solid blue), and Mn$_{1.6}$PtSn$_{0.9}$ (solid black). Positive (negative) DOS are the majority (minority) projected states. Inset: Enlarged area around the Fermi energy. (b) The band-structure of \mps{2} around the Fermi energy. (c) The AHE calculated as a function of the Fermi energy for \mps{2} (grey dashed) and \mps{1.5} (solid red). The points show the experimental anomalous Hall conductivities. (d) Calculated $M(H)$ loops for \mps{1.5}~with the out of plane magnetic field above and below the spin reorientation. The insets show the magnetization for in plane magnetic field. (e)~\dmdh~derived from the $M(H)$ loops.}
	\label{fig:F3}
\end{figure}

\subsection*{First principle calculations}	
We carry out first-principle calculations on \mps{1.5}~within the FLAPW method using the FLEUR code~\cite{FLEUR}. Here, we find that the moments on the Mn$_{\mathrm{I}}$ and Mn$_{\mathrm{II}}$ sublattice are 3.72 and 3.4~\textit{$\mu$}$_{\mathrm B}$, respectively~\cite{Vir2019a}. Due to the hybridization, there is a small moment induced on the Pt and Sn sites of 0.3 and 0.1~\textit{$\mu$}$_{\mathrm B}$, respectively. We allow the directions of the moments to relax with SOC and find the angle between the Mn$_{\mathrm{I}}$ and Mn$_{\mathrm{II}}$ sublattices to be 52~$\degree$. This shows that the ground state is a noncoplanar spin structure. The collinear ferrimagnetic (FiM) structure has an energy that is 60~meV larger while the collinear ferromagnetic state is 140~meV larger.

Upon substitution of  Mn onto a Pt-8$c$ (Sn-8$d$) site, the symmetry of the system is lowered, and the average canting angle of the Mn sublattices is decreased (increased) to 47~$\degree$ (56~$\degree$). In Fig.~\ref{fig:F3}a, we plot the spin decomposed density of states (DOS) for the ferrimagnetic (dashed), ferromagnetic (solid black), a single Pt substituted by Mn, and a single Sn substituted by Mn per unit cell. The positive values are for the majority spins, and the negative for the minority spins. For the Mn substituted onto a Pt or Sn site, the Mn prefers to align antiferromagnetically to the net moment, thereby decreasing the total moment. For the substitution at the Sn site the DOS at the Fermi energy decreases. $\Delta_l$ denotes the exchange field generated by each Mn sublattice ($l$). In Fig.~\ref{fig:F3}b we plot the bands of \mps{2}, which are significantly less complicated than that of \mps{1.5} (see Fig.~S7). We calculate the anomalous Hall conductivity (AHC) for the collinear ferromagnetic structures where we assume the external magnetic field has aligned the moments of Mn\textsubscript{II}-8$d$ and Mn\textsubscript{I}-4$a$ along the $a$-axis. In the \mps{2} we find an AHC of 95~S/cm (273~S/cm) along the $b$-axis ($c$-axis) and in \mps{1.5} we find an AHC of 155~S/cm (232~S/cm) along the $b$-axis ($c$-axis). Along with that, we plot the AHC as a function of energy for \mps{2}~and \mps{1.5}~ for the collinear magnetic spin texture along with the experimental anomalous Hall conductivities (Fig.~\ref{fig:F3}c). 

\subsection*{Atomistic spin calculations}	
The origin of the noncollinear magnetic textures is due to a competition of classical exchange interactions between the inter-site interactions of Mn$_{\mathrm{II}}$-8$d$ and Mn$_{\mathrm{I}}$-4$a$ and the intra-site interactions. This can be modeled by a classical exchange Hamiltonian
${\cal H}=-J_1\sum_{<ij>,l,l'}({\mathbf S}_{i}^{l}\cdot {\mathbf S}_{j}^{l'})+-J_2\sum_{<ij>}{\mathbf S}_{i}^{\textrm I}\cdot {\mathbf S}_{j}^{\textrm I}+-J_3\sum_{<ij>}{\mathbf S}_{i}^{\textrm{II}}\cdot {\mathbf S}_{j}^{\textrm{II}}$. Here, $S$ is the unit vector of spin for each sublattice ($l$), $J_1$ is the inter sublattice interaction, $J_2$ and $J_3$ are the Mn$_{\textrm{II}}$-8$d$ and Mn$_{\textrm{I}}$-4$a$ intra-site interactions for each sublattice, respectively. All interactions together are necessary for a noncollinear spin texture. The positive Mn$_{\mathrm{II}}$-8$d$ interactions, $J_2$, align the spins parallel and are the largest contribution to the ordering temperature. The Mn$_{\mathrm{I}}$-4$a$ interactions, $J_3$, are negative, which favors an antiparallel alignment. Lastly, the inter-site interaction, $J_1$, is positive and causes a noncoplanar magnetic structure depending on the relation with the other interactions. When $J_1$ is weak ($J_1<\frac{1}{4}J_2$), the coupling between Mn$_{\mathrm{II}}$-8$d$ and Mn$_{\mathrm{I}}$-4$a$ sublattices is weak, and the noncoplanar structure stabilizes as the ground state. In the noncoplanar spin texture there is a finite chiral product, $\chi={\mathbf S}_{i}\cdot({\mathbf S}_{j}\times{\mathbf S}_{k})$, that gives rise to the real space Berry curvature. When $J_1$ is strengthened, the Mn$_{\mathrm{I}}$-4$a$ sublattice follows that of the Mn$_{\mathrm{II}}$-8$d$ for a collinear alignment and is therefore the origin of the spin reorientation. We map these interactions on the atomistic Landau-Lifshitz-Gilbert equation: 
	
\begin{equation}\label{eq:LLG}
\centering
	\frac{\partial {\mathbf S}_i}{\partial t}=-\frac{\gamma}{(1+\lambda^2)}\left[{\mathbf S}_i\times{\mathbf H}_{\rm eff}^{i}+\lambda{\mathbf S}_i\times({\mathbf S}_i\times{\mathbf H}_{\rm eff}^{i})\right].
\end{equation}
	
Here, ${\mathbf S}_i$ is the unit vector of spin at site $i$, ${\gamma}$ is the gyromagnetic ratio and $\lambda$ the microscopic damping. ${\mathbf H}_{\rm eff}^{i}=-\frac{1}{\mu_s}\frac{\partial {\mathcal H}}{\partial {\mathbf S}_{i}}+{\mathbf H}_{\rm th}^{i}$ is the effective field with thermodynamic fluctuations (${\mathbf H}_{\rm th}^{i}$) modeled by Langevin dynamics and ${\mathcal H}$, which includes the exchange interactions, a uniaxial anisotropy and the external magnetic field. Within the vampire code, we simulate a 15x20x15~nm \mps{1.5} for periodic boundary conditions in the $b$ and $c$ directions and an open boundary in the $a$ crystal axis. In Fig.~\ref{fig:F3}d we plot the $M(H)$ loops, below and above the spin reorientation for a magnetic field along the $a$-axis. The inset shows the $M(H)$ loops for the field along the $c$-axis. Fig.~\ref{fig:F3}e shows the \dmdh~curves of the theoretical calculations. Similar to the experimental curves, the theoretical \dmdh~shows the transition field, \hs (open square), of the magnetic texture. Below \hs~the magnetic textures induces a change in the Berry curvature, thereby adding a chiral product ($\chi$) contribution to the Hall effect.

\begin{figure}[h!]
	\centering
	\includegraphics[width=\columnwidth]{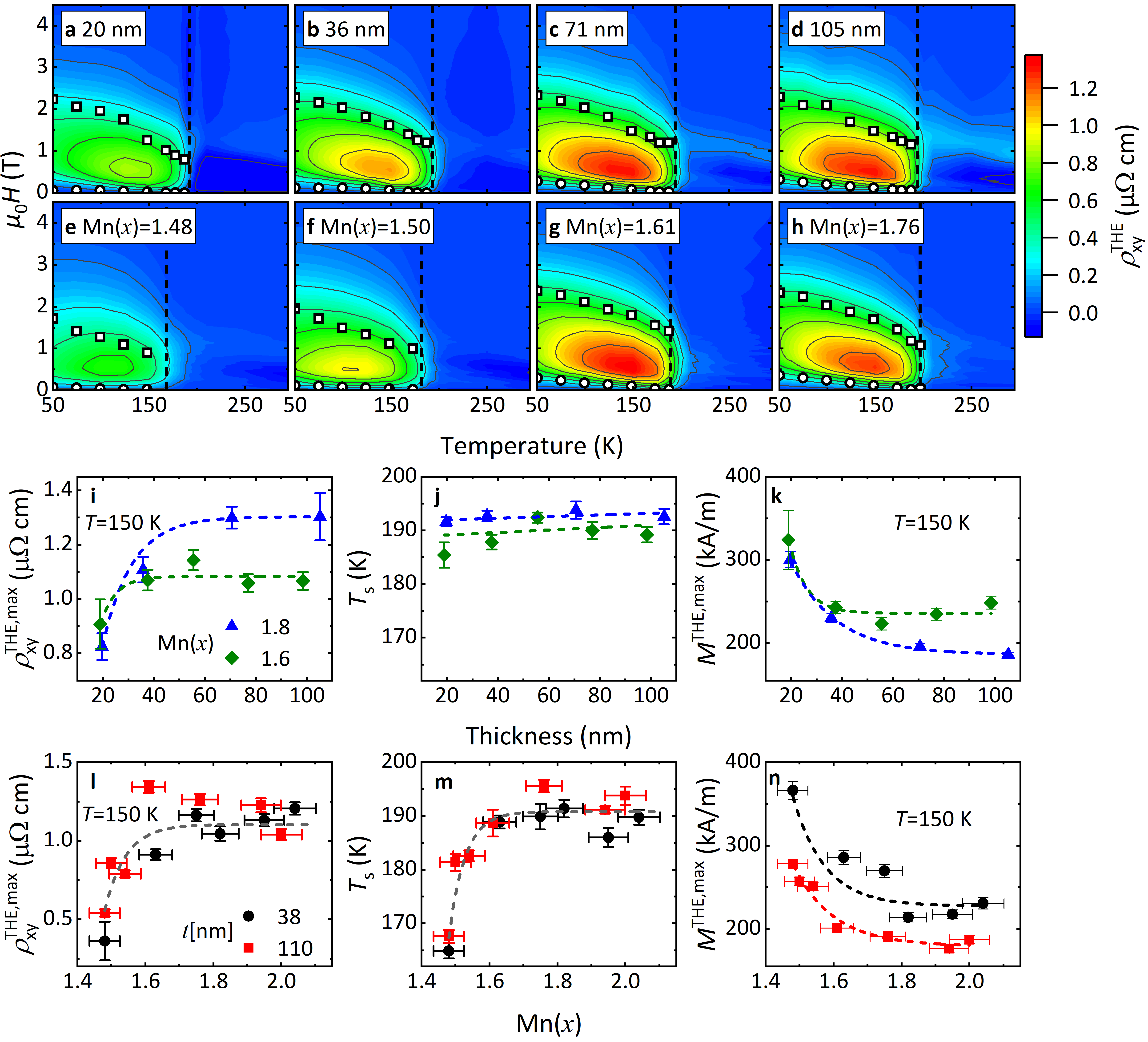}\\
	\caption{\rt~as a function of temperature and field for (a)-(d) \mps{1.8} films with multiple thicknesses and (e)-(h) 38~nm thick \mps{x}~films with multiple \concx. The overlays are the first-order transition of the noncoplanar texture orientation~(circle) and second-order transition into the collinear alignment~(square) extracted from \dmdh. \ts~is indicated by a dashed line. (i)-(k) \rtm, \ts~and $M^{\mathrm{THE,max}}$~as a function of film thickness. (l)-(n) \rtm, \ts~and $M^{\mathrm{THE,max}}$~as a function of \concx. Dashed lines are guides for the eye. \rt~and $M^{\mathrm{THE,max}}$~were extracted from data recorded at 150~K.}
	\label{fig:F5}
\end{figure}

\subsection*{Chiral-type Hall effect from magnetic texture}	
In Fig.~\ref{fig:F5}a-h we combine the results of the extracted \rt~and the transition fields of \dmdh~as a function of temperature and field for the \conc{1.8} thickness series (top row) and the $t=$110~nm \concx~series (bottom row). The dashed line indicates \ts, the squares indicate the second-order transition  and the circles indicate the first-order transition. For both (\conc{1.6} and \conc{1.8}) thickness series, we plot the maximum topological Hall resistivity (\rtm), the magnetization at the corresponding field of \rtm~and \ts~in Fig.~\ref{fig:F5}i-k, respectively. The dashed lines are asymptotic fits of the experimental data. As can be expected, \ts~shows little variation with thickness (Fig.~\ref{fig:F5}j). Further, \rtm~saturates above 40~nm ((Fig.~\ref{fig:F5}i). When compared with the magnetization dependence in Fig.~\ref{fig:F5}k, the origin of this mechanism can be related to the magnetic dipole-dipole interaction, where the magnetization increases with a decrease in film thickness~\cite{Srivastava2020}. 
	
In Fig.~\ref{fig:F5}l-n, we show the dependence of \rtm, the magnetization at the corresponding field of \rtm~and \ts~as a function of \concx~for $t$=38~nm (black) and 110~nm (red) films. The saturating evolution of \rtm~as function of \concx~is distinct from the thickness dependence of \rtm, mainly determined by anisotropies. Here, \rtm~is determined by the variation of exchange constants as a function of \concx~that in turn determine the size of the chiral product. The magnetization in Fig.~\ref{fig:F5}n shows a decrease with \concx, while \ts~(Fig.~\ref{fig:F5}m) increases, both saturating above \conc{1.75}. This indicates an increase of the chiral product with \concx. The increase is related to a higher Mn$_{\mathrm{II}}$ sublattice occupation that prefers an antiparallel alignment of the magnetic moments due to the antiferromagnetic exchange, in turn increasing the overall canting angle (see Fig.~\ref{fig:F4}b). Furthermore, due to only slight changes in the electronic structure and Berry curvature (Fig.~\ref{fig:F3}a), the origin of the variation in the \rtm~is due to the described canting of the spins in the magnetic field. 
	
\begin{figure}[h!]
	\centering
	\includegraphics[width=\columnwidth]{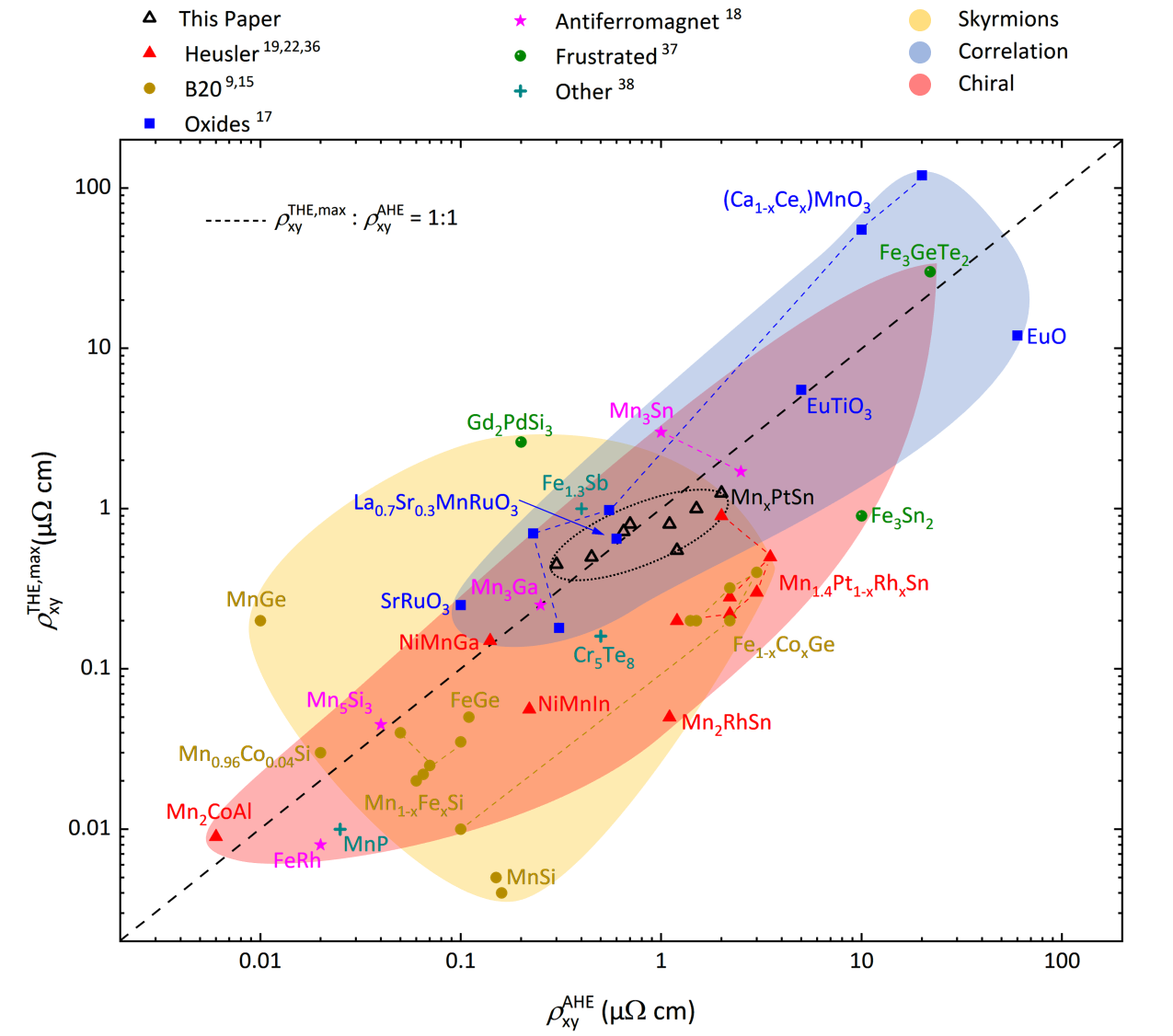}\\
	\caption{Comparison of \rtm~as a function of \ra~of our \mps{x} thin films (black open triangles) and other previously reported relevant materials. The colored shaded areas indicate the three main classes: skyrmion (yellow), chiral (red) and correlated (blue). The dashed line is a guide to the eye of a 1:1 scaling.}
	\label{fig:F6}
\end{figure}
	
In Fig.~\ref{fig:F6}, we show the \ra~versus \rtm~on a logarithmic scale for three classes of materials: skyrmion, chiral and correlated. Within these classes, we select several relevant systems. The Heusler compounds, known to have chiral spin textures and to host antiskyrmions, Mn$_{1.4}$Pt$_{1-x}$Rh$_{x}$Sn, Mn$_{2}$RhSn, Mn$_{2}$CoAl, NiMnGa and NiMnIn~\cite{Vir2019a,Kumar2020,Rana2016,*Ludbrook2017,*Zhang2019b}. The well known B20 skyrmion systems, MnGe, MnSi, Mn$_{1-x}$Fe$_{x}$Si, Fe$_{1-x}$Co$_{x}$Ge and Mn$_{0.96}$Co$_{0.04}$Si~\cite{Neubauer2009,Kanazawa2011,*Huang2012,*Franz2014,*Spencer2018,*Yokouchi2014}. The oxide materials, with many being correlated and showing magnetic domains/bubbles, EuO, EuTiO$_3$, (Ca$_{1-x}$Ce$_{x}$)MnO$_3$, SrRuO$_3$ and La$_{0.7}$Sr$_{0.3}$MnRuO$_3$~\cite{Ohuchi2015,*Takahashi2018,*Vistoli2018,*Nakamura2018,*Qin2019}. Lastly, antiferromagnetic compounds~\cite{Suergers2014,*Liu2017,*Yan2019,*Rout2019,*Taylor2019}, frustrated systems~\cite{Kurumaji2019,*Wang2017,*Li2019}, and systems that we classify as others~\cite{Wang2019,*Shiomi2012a,*Shiomi2012b}. Interestingly, Mn is one of the most frequented elements in these compounds. Furthermore, on a logarithmic plot, all these compounds straddle the one to one ratio line for \ra~and \rtm. The correlated materials comparatively show both large \ra~and large \rtm. Whereas, the skyrmion materials show the most substantial variation away from the one to one ratio line. Lastly, the chiral compounds show a significant variation across the entire graph. 
	
Skyrmionic textures are nominally considered to be long-range textures that give rise to the THE, due to the emergent field of the skyrmion lattice~\cite{Bruno2004}. This is usually pictured as an electron that couples adiabatically to a smoothly varying magnetization lattice that can be several orders of magnitude larger than the crystal unit cell. Whereas, in correlated systems the density of free carriers can be reduced orders of magnitude to the comparative systems~\cite{Ohuchi2015,*Takahashi2018,*Vistoli2018,*Nakamura2018,*Qin2019}. Here, the coupling of the conduction electron to the magnetic sublattice is enhanced and not only the Berry phase (adiabatic) but also non-adiabatic processes should be taken into consideration~\cite{Denisov2017,Onoda2004}. The chiral magnetic structure suffices in both regimes. The overall strength of the THE depends on how strongly the conduction electrons couple to the magnetic sublattice, wherein these systems the magnetic lattice is on the order of the crystal unit cell~\cite{Denisov2017,Onoda2004}. This results in two effects: i) the rapid variation in the sublattice leads to non-adiabatic effects; ii) Berry phases in both real and momentum space emerge. Specifically, \mps{x} shows clearly that as the canting of the spins increases, the THE is also enhanced, and the non-adiabatic mechanisms become crucial. 
	
Nominally, the exciting contribution to the AHE is the intrinsic adiabatic Berry curvature~\cite{Nagaosa2010}. The requirement for the AHE is the lack of time-reversal symmetry and the splitting of orbital degeneracy. In ferromagnets, this is realized by a finite magnetization and the SOC. Whereas, in the noncollinear antiferromagnets the time-reversal symmetry is broken by the all in or all out structure. The orbital degeneracy is lifted by the conduction spin coupling to the magnetic lattice. Likewise, for the THE the time-reversal symmetry is broken by the direction of the cone angle of the noncoplanar spins. The SOC is replaced by the relative change in the direction of the magnetization in the magnetic sublattice~\cite{Zelezny2017}. This leads to very similar but distinct origins for the AHE and THE. While the overall magnitude of the effects depends on the intimate nature of the electronic structure, the values of the two will always be comparable when sharing the same electronic structure. An exception may occur when there is a topological Fermi surface where the THE has a weaker dependence on the Fermi sea contribution in the weak coupling regime~\cite{Denisov2018}. 
	
In summary, \mps{x}~and the host of compounds in this Heusler family, are of particular interest because the fundamental exchange parameters are tunable by chemical substitution~\cite{Graf2011,Kumar2020}. Specifically, we show that the role of \concx~is to tune the ratio of antiferromagnetic and ferromagnetic exchange, which is crucial in the stabilization of antiskyrmions. Fine-tuning of \concx~would allow for discrete adjustments of the antiskyrmions below the spin reorientation transition temperature. Furthermore, we find that this effect should be observable from low temperatures up to $\sim$200~K depending on \concx. As expected, the magnetic dipole-dipole interactions play the most significant role in the thin film limit. The most substantial change in the THE is due to the \concx, which tunes the relation of the exchange couplings and thereby the topological Berry curvature. The increase of \concx~increases the unit cell volume, therefore we can infer that the noncoplanar spin texture is also coupled to the lattice size, and can be tuned as in multiferroic heterostructures. 
	
	
\bibliography{MnPtSn_Exchange_2020}
	
	
\section*{Acknowledgement}
	
The authors acknowledge funding by the Deutsche Forschungsgemeinschaft (DFG, German Research Foundation) under SPP 2137 (Project No. 403502666) and the European Union’s Horizon 2020 research and innovation programme, under FET-Proactive Grant agreement No. 824123 (SKYTOP). D.K acknowledges the Ministry of Education of the Czech Republic Grant No. LM2018110 and LNSM-LNSpin, and the Czech Science Foundation Grant No. 19-28375X. HAXPES measurements were performed at beamline P22 of PETRA~III (Proposal No.~I-20181156). We thank A. Gloskovskii (PETRA, Hamburg), B. Balke (Fraunhofer IWKS, Hanau) and W. Xie (TU Darmstadt) for help with the HAXPES experiments. Funding for the HAXPES instrument at beamline P22 by the Federal Ministry of Education and Research (BMBF) under contracts 05KS7UM1 and 05K10UMA is gratefully acknowledged.

\end{document}